\documentclass[aps,amsmath,amssymb,twocolumn,superscriptaddress,]{revtex4}%
\usepackage{amsmath}
\usepackage{xcolor}
\usepackage{graphicx}
\usepackage{dcolumn}
\usepackage{bm}
\usepackage{float}
\usepackage{epsfig}
\usepackage{amsfonts}%
\usepackage{color}
\usepackage[utf8]{inputenc}
\usepackage{amssymb}
\usepackage{natbib}
\usepackage{soul}
\usepackage{scalerel}
\usepackage{cancel}
\usepackage{appendix}

\begin{document}

\preprint{APS/123-QED}

\title{Adiabatic and Deterministic Routes to Soliton Combs in Non-Hermitian Kerr Cavities}

\author{Salim B. Ivars}
\email{salim.benadouda@icfo.eu}
\affiliation{ICFO - Institut de Ciències Fotòniques, The Barcelona Institute of Science and Technology, 08860 Castelldefels (Barcelona), Spain}

\author{David Artigas}
\affiliation{ICFO - Institut de Ciències Fotòniques, The Barcelona Institute of Science and Technology, 08860 Castelldefels (Barcelona), Spain}

\affiliation{Department of Signal Theory and Communications, Universitat Politècnica de Catalunya, 08034 Barcelona, Spain}

\author{Carlos Mas Arabí}
\author{Carles Mili\'{a}n}
\affiliation{Institut Universitari de Matem\`{a}tica Pura i Aplicada, Universitat Polit\`{e}cnica de Val\`{e}ncia, 46022 Val\`{e}ncia, Spain}



\begin{abstract}
We present a cardinal solution for the long-standing and fundamental problem associated with the adiabatic, reversible, and controlled excitation of both dark and bright solitons in Kerr micro-resonators with normal group velocity dispersion. Our findings stem from the inclusion of a localised non-Hermitian potential, which we use to drastically reshape the characteristic collapsed snaking structure associated with such solitons. Consequently, we demonstrate a novel snaking-free bifurcation landscape where solitons of all possible widths are continuously connected via the dynamic change of the cavity detuning, and hence dissipative localised states of unprecedentedly high pump-to-comb conversion efficiencies can be excited in an adiabatic, deterministic, and reversible fashion. Our fundamental discovery has practical implications of paramount importance for frequency comb generation in all-normal dispersion cavities, which are key to comb generation in most spectral regions away from the telecom bands.
\end{abstract}

\maketitle
\begin{figure*}
    \centering
    \includegraphics[width=1\textwidth]{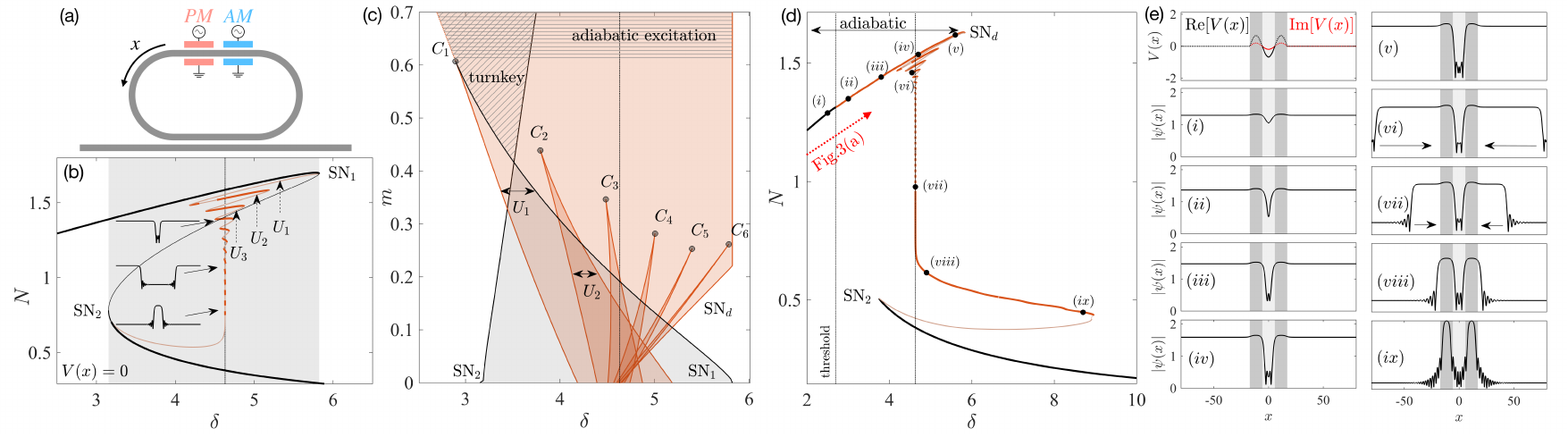}
     \vspace{-20pt}
    \caption{
    (a) Sketch of the optical Kerr cavity with an integrated non-Hermitian \textcolor{black}{phase (PM) and amplitude (AM) modulation}. (b) The typical bifurcation diagram of single LSs without potential (orange curve) is embedded within the reference nonlinear resonance formed by the homogeneous steady states (black line). Saddle nodes SN$_{1,2}$ are marked for reference. Labels $U_n$ ($n=1,2,...$) mark LSs unstable branches. \textcolor{black}{Stable (unstable) states are denoted by thick (thin) lines in all bifurcation diagrams}. (c) Position of the key SNs as a function of the potential strength, $m$, and detuning, $\delta$. $C_n$ (black dots) mark the cusps associated with the disappearance of the unstable branches $U_n$. Orange areas delimit regions where stable LSs are found. Patterned areas with tilted and horizontal lines mark regions where LSs can be excited in a turnkey and adiabatic fashion, respectively. \textcolor{black}{SN$_d$ is associated to the widest soliton continuously accessible when $C_1$ is suppressed}. Grey areas in (b),(c) mark the homogeneous state bistability region. Vertical lines in (b-d) mark the Maxwell point (MP). (d) Bifurcation diagram for $m=0.7$ and $\phi=0.91\pi$ showing LSs (orange) emerging from homogeneous states (black). (e) Panels showing the potential profile (top-left) and selected LSs. Light (dark) gray regions correspond to high (low) losses. \textcolor{black}{$q=0.275$ in all the figure.}}
    \label{fig:1}
\end{figure*}

%
The first observations of optical Kerr microcombs \cite{del2007optical,del2012hybrid,li2012low} and temporal solitons in passive Kerr resonators \cite{leo2010temporal,herr2014temporal,xue2015mode} had a profound impact on the technological developments of frequency comb-related applications \cite{cundiff2003colloquium,herr2016dissipative,pasquazi2018micro,gaeta2019photonic} as well as in the fundamental advances in the field of dissipative soliton physics \cite{akhmediev2008dissipative,purwins2010dissipative,yang2017stokes,ivars2023snakes}.
 In particular, solitons in microring resonators featuring normal group velocity dispersion (GVD) \cite{xue2015mode,xue2015normal} were recognised as essential frequency comb carriers for applications requiring high pump-to-comb efficiencies \cite{lobanov2015frequency, xue2017microresonator,yang2024efficient} or operation points where anomalous GVD is difficult or impossible to obtain \cite{xue2016normal,okamoto2021fundamentals,rebolledo2023platicon}. The important above attributes \textcolor{black}{contrast with the challenge of experimentally exciting modulationally unstable states in this regime \cite{parra2016dark}. Different strategies have proven successful for entering the bistable area \cite{xue2015mode,jang2016dynamics,kim2019turn,xue2019super,nazemosadat2021switching,rebolledo2023platicon,li2025universal,sanyal2025nonlinear} as well as for turnkey generation \cite{lobanov2019generation,lobanov2015generation} or dynamical switching to narrow dark solitons \cite{nazemosadat2021switching}}. Despite this remarkable progress, the deterministic \textcolor{black}{and continuous} excitation \textcolor{black}{of the coveted bright soliton (and associated highly efficient combs)} in the deep normal GVD region represents a milestone yet to be achieved since such a possibility would enable an unprecedented simple access to frequency combs of very high pump-to-comb efficiency, essential for applications such as communications \cite{marin2017microresonator,corcoran2025optical}, lidar \cite{trocha2018ultrafast} or spectroscopy \cite{picque2019frequency}.

From a more fundamental standpoint, dissipative solitons in the normal GVD regime of a Kerr cavity belong to a general class of localised states (LSs) arising via locking of heteroclinic orbits in systems with multiple stable coexisting homogeneous steady states. They indeed exist in a plethora of nonlinear models related to the paradigmatic Swift–Hohenberg \cite{dawes2008localized,avitabile2010snake}, Boussinesq \cite{knobloch2005homoclinic}, Navier-Stokes \cite{tseluiko2014collapsed}, Ginzburg Landau \cite{yochelis2006reciprocal}, or Schrödinger \cite{yulin2010discrete,parra2016dark} equations, amongst others \cite{seidel2022normal} \textcolor{black}{- originating in very different physical contexts}. A strong common feature of all such LSs is that, because of their front-locking nature, they exhibit a \textit{collapsed} snaking structure in the parameter space, as predicted by fundamental theorems \cite{knobloch2005homoclinic}, which strongly dictates the possible access routes to those states. 
In addition, the huge amount of different multi-stable solitons existing under the above paradigm \cite{parra2016dark,talla2017existence}, makes it inconceivable to aim at a deterministic and selective excitation of such solitons and the associated combs in a simple way. While multi-stability of LSs is often sought and induced to enrich the system's solutions in many areas of nonlinear sciences \cite{fang2022multistability}, in practice, complex strategies are often required to excite the desired LSs, such as the annihilation of competing attractors or the introduction of cumbersome stochastic perturbations (see, e.g. Ref. \cite{pisarchik2014control} for a review in the topic). 

In this Letter, we introduce a very unique mechanism to drastically unfold \textcolor{black}{and reshape} the multi-stable bifurcation structure associated with LSs with normal GVD so that all desired stable LSs, namely dark and bright solitons of different widths, \textcolor{black}{emerge smoothly and continuously from the homogeneous state, being} distributed over the same stable branch in the parameter space. \textcolor{black}{This opens} the possibility to smoothly \textcolor{black}{excite and transform} them at will in a reversible\textcolor{black}{, adiabatic  and deterministic} fashion. This unprecedented situation is achieved by introducing a non-Hermitian localised potential which, in turn, induces a collection of dislocated cusp bifurcations which are inherently associated with dramatic qualitative changes in the behaviour of nonlinear systems \cite{kuznetsov1998elements}. We emphasise that, while cusp bifurcations are commonly regarded as significant because they induce bistability, we exploit them in the opposite sense, using them to remove multi-stability while keeping the nature of the system's LSs.

\textcolor{black}{The design and implementation of complex potentials have been explored in a variety of systems, including acoustics \cite{huang2024acoustic}, Bose-Einstein condensates \cite{barontini2013controlling} and electronic circuits \cite{huerta2023generating}, among others (see, e.g. \cite{yu2025comprehensive} for a review). In the context of optics and frequency combs,} recent technological advancements have demonstrated integrated modulators in the Hermitian \cite{ren2019integrated,henke2021integrated} and non-Hermitian \cite{sinatkas2021electro,hu2025integrated,wang2025dynamic} cases in a wide variety of geometries. \textcolor{black}{The introduction of} electro-optic modulators (EOMs) is bringing novel features such as combs with high tunable frequency spacing \cite{zhang2019broadband} and efficiency \cite{hu2022high}, as well as relevant physical predictions for synthetic frequencies \cite{dutt2020single,yuan2021synthetic,tusnin2020nonlinear} or the hybridisation of soliton formation paradigms \cite{ivars2024hybrid}.



We assume the temporal dynamics of the intra-cavity electric field envelope, $\psi$, is well described by the normalised damped-driven nonlinear Schrödinger equation \cite{chembo2013spatiotemporal,lugiato1987spatial,haelterman1992dissipative} generalised to include refractive index modulations (potential) \cite{tusnin2020nonlinear}:
\begin{align}
\label{eq:1}
\partial_t \psi &= -i \partial_x^2 \psi - (1 + i\delta) \psi + 2i |\psi|^2 \psi + V(x) \psi + h, \\
V(x) &= m e^{-i\phi} \cos(qx)\, \Theta\left({3\pi}/{2q} - |x|\right) \tag{2} \label{eq:potential}.
\end{align}
The terms on the right-hand side of Eq.\ref{eq:1}, respectively, account for normal GVD, optical losses, cavity-laser detuning, Kerr nonlinearity, complex potential, and coherent driving strength. The potential defined in Eq.\ref{eq:potential} corresponds to a non-Hermitian type of travelling wave modulation \cite{yariv1983optical}: $m$ represents its depth, $q$ is the spatial frequency, $\phi$ is a phase controlling the real and imaginary components, and $\Theta$ is the Heaviside step function. In practice, such non-Hermitian potentials can be introduced in micro-cavities via integrated phase and amplitude EOMs \cite{hu2025integrated}, as sketched in Fig.\ref{fig:1}(a).


The key effect in our results is the remoulding of the typical bifurcation diagram associated with solitons in the normal GVD, represented in Fig.\ref{fig:1}(b). There, LSs, formed via pairs of bounded switching waves \cite{parra2016dark}, display a complex snaking diagram that collapses to the so-called Maxwell Point (MP) due to the absence of a pinning potential on their low amplitude oscillatory tails when the fronts move apart from each other \cite{knobloch2005homoclinic}. Associated with these oscillations, stable states with different widths are located within different branches [see insets] separated by intrinsically unstable ones [$U_1$, $U_2$, ...] delimited by Saddle Node (SN) bifurcations. Such collapsed diagram features a vast collection of multi-stable LSs, evidencing the difficulty to neatly excite LSs of different families (different widths) in experiments \cite{pisarchik2014control}, particularly those of bright nature (at low norm, $N\equiv\int|\psi|dx$) associated with highly efficient combs \cite{xue2017microresonator}.

Our primary goal is to drastically reshape the snaking diagram in Fig.\ref{fig:1}(b) by eliminating its nested collapsed nature and the associated unstable branches $U_1$, $U_2$, etc., without frustrating the existence of the LSs therein. To this end, we recall that collapsed snaking arises as a consequence of the absence of a strong pinning potential \cite{knobloch2005homoclinic}\textcolor{black}{, which stimulates the idea that the introduction of a potential may counteract the system's trend to the collapsed snaking}. However, the introduction of Hermitian potentials via driving \cite{lobanov2015generation} or intra-cavity \cite{sun2023multimode} modulations produces an enhancement of the unstable branch $U_1$, contrary to our aims. Hence, the natural idea of introducing a non-Hermitian potential arises \textcolor{black}{and, as we show below, turns out to be most successful}. Indeed, as represented in Fig.\ref{fig:1}(c), upon the increase of the potential depth, the unstable branches ($U_n$) cease to exist as the SNs delimiting them collapse into a series of cusps, $C_n$. Crucially, the arrangement of cusps in the $\{\delta,m\}$ plane result in the bifurcation diagram in Fig.\ref{fig:1}(d), for $m=0.7$, where various top (in norm) branches associated to narrow dark solitons in Fig.\ref{fig:1}(b) are now aligned in detuning and continuously connected without intermediate unstable regions, and hence, such LSs [see panels $(i)-(v)$] can be continuously excited in an adiabatic fashion from the homogeneous single frequency states simply by red-detuning the driving laser frequency. \textcolor{black}{This comes as a result of the unfolding of the first six cusps.} In addition, LSs can be changed in shape in a reversible fashion via red or blue detuning of the pump. This adiabatic and reversible excitation mechanism, possible above the cusp $C_1$ ($m\gtrsim0.61$), illustrates our central fundamental result. 

The above effect reaches a natural limit when dark solitons acquire a width comparable to the confining potential [panel $(v)$]\textcolor{black}{. Accordingly, $q$ determines the widest state continuously connected to the homogeneous state. Then, the} LSs undergo a SN bifurcation characterised by the formation of a secondary dark soliton in the regions of the cavity unaffected by $V(x)$ [see panels $(vi)-(vii)$], \textit{recovering} the collapsed snaking structure. Eventually, when the secondary state is wide enough to feel $V(x)$, the LS is transformed into a double bright soliton [see panels $(viii)-(ix)$] exhibiting a remarkably broad existence and stability region. The latter bright LSs do not exist when $V(x)$ is composed solely of its central layer [cf. top panel in Fig. \ref{fig:1}(e)], which illustrates the system's richness and potential interest in pursuing further engineering of $V(x)$. It is also important to realise that LSs exhibit a turnkey excitation window ($m\gtrsim 0.42$) where the dark solitons are the only stable solution of the system, a property also observed recently \cite{shen2020integrated,rowley2022self,jin2021hertz} and predicted in other types of LSs and patterns in several types of damped-driven Nonlinear Schr\"{o}dinger equations \cite{sun2023robust,ivars2023snakes}.


\begin{figure}
    \centering
    \includegraphics[width=\columnwidth]{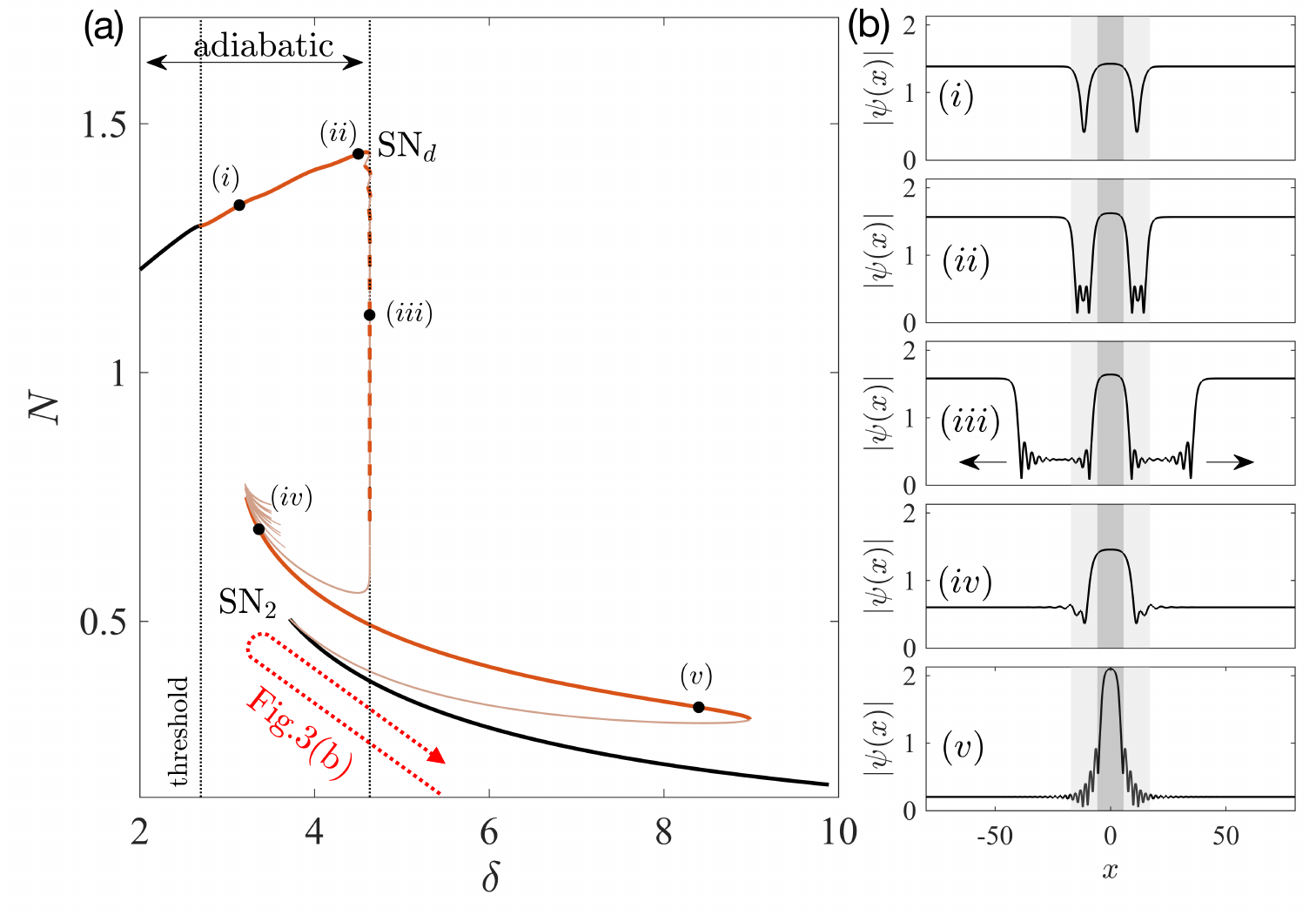}
     \vspace{-20pt}
    \caption{
    (a) Bifurcation diagram displaying the connection between the extended quasi-homogeneous (black) and localised (orange) states for $m=0.7$, \textcolor{black}{$\phi=-0.09\pi$, $q=0.275$}. (b) Panels showing the potential with the same colour code as in Fig. \ref{fig:1}(e), and selected LS solutions located near the labels $(i)-(v)$ on the bifurcation diagram. 
    }
    \label{fig:2}
\end{figure}

\begin{figure}
    \centering
    \includegraphics[width=\columnwidth]{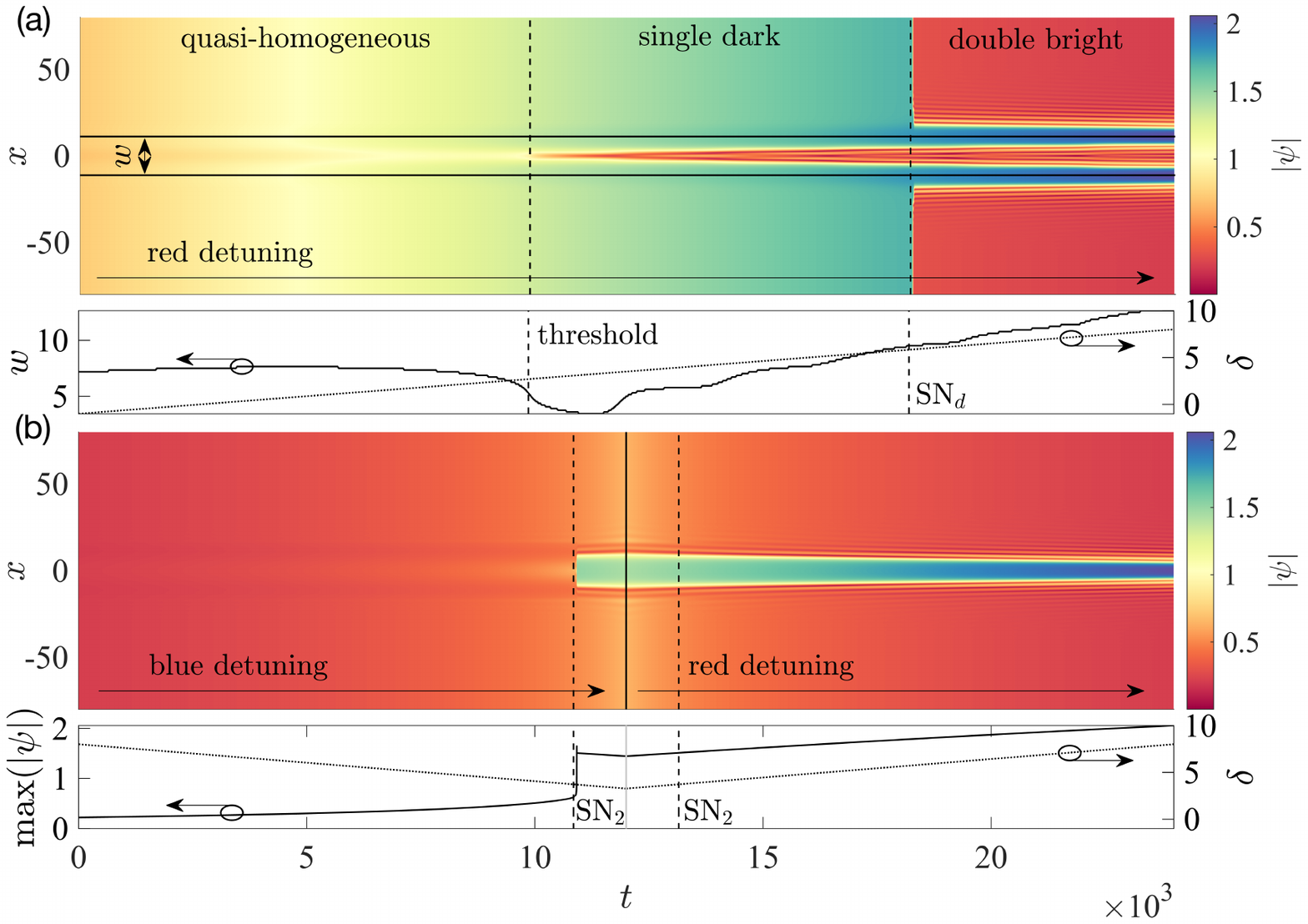}
     \vspace{-20pt}
    \caption{Real-time excitation and manipulation of frequency combs via cavity detuning. (a) Single dark soliton smoothly emerging from the cavity background state under detuning red-shift and its subsequent decay to the pair of bright LSs corresponding to Fig. \ref{fig:1}(c). Dashed lines denote the localisation of the adiabaticity threshold (left) and the SN$_d$ associated with the abrupt switching (right). Bottom panel corresponds to the temporal evolution of $\delta$ and the FWHm, defined for the field in between the two horizontal black lines. (b) Excitation of single bright LS via abrupt switching for a blue-shift in detuning and subsequent smooth manipulation via detuning red-shift corresponding to Fig. \ref{fig:2}. Dashed vertical lines denote the position of SN$_2$. The solid line marks the time at which we change the direction of the detuning sweep. The bottom panel corresponds to the temporal evolution of cavity detuning and maximum amplitude of the field. 
    }
    \label{fig:3}
\end{figure}
%

\begin{figure*}
    \centering
\includegraphics[width=\textwidth]{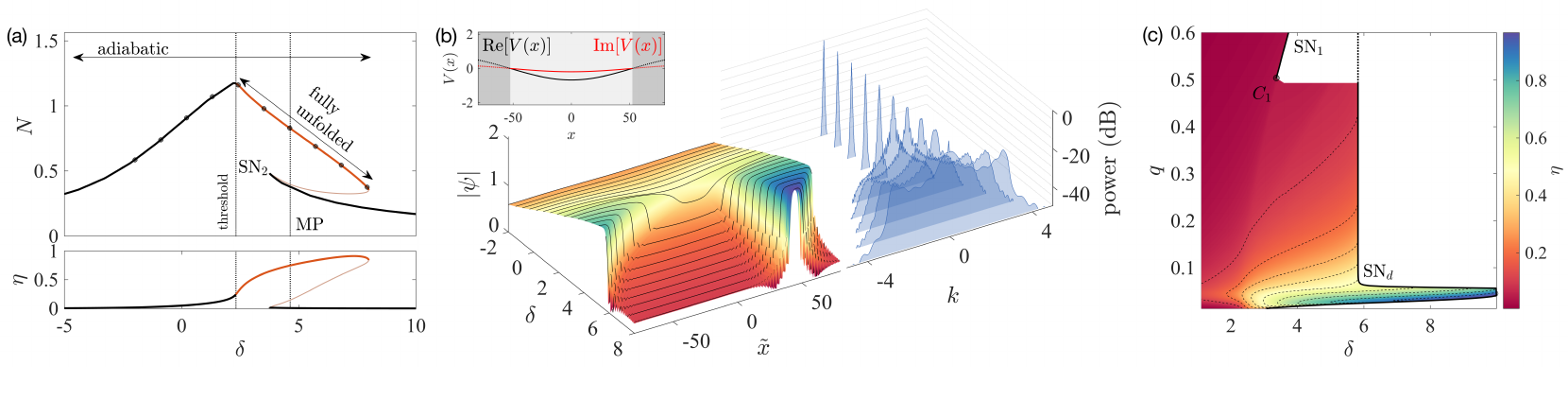}
     \vspace{-20pt}
    \caption{(a) [top] Fully \textcolor{black}{unfolded} bifurcation diagram for $q=0.03$ and $\phi=0.91\pi$ with all possible stable dark and bright states located on the stable orange branch. [bottom] Pump-to-comb efficiency estimated from line contrast in the intra-cavity field. (b) [left] Field profiles vs $\delta$ for stable solitons in (a) and [right] selected corresponding frequency combs \textcolor{black}{where $k$ is the wavenumber}. Black dots in (a) correspond to the combs in (b). The spatial coordinate $\tilde{x}$ is conveniently shifted from $x$ in our model to better visualise the soliton profiles, which are all centred at $x=0$, which represents the minimum of the potential (top inset).  
    (c) Line contrast $\eta$ for the states continuously emerging from the CW in the $(\delta,q)$ plane.}
    \label{fig:4}
\end{figure*}

Exhaustive numerical modelling reveals two qualitatively different bifurcation diagrams when scanning the $\{m,\phi\}$ plane, i.e., the complex amplitude of our non-Hermitian potential $V(x)$: one type occurs for $\phi\in(\pi/2,3\pi/2)$, discussed in Fig.\ref{fig:1}(d), and the other one for $\phi\in(-\pi/2,\pi/2)$. The latter type, illustrated in Fig.\ref{fig:2} for $m=0.7$, \textcolor{black}{$\phi=-0.09\pi$ and $q=0.275$}, produces a unfolding of the snaking structure \textcolor{black}{associated to the first four solitons} analogous to that in Fig.\ref{fig:1}(d), but now associated with a pair of dark solitons [panels $(i)-(ii)$ in Fig.2(b)]. Due to a transcritical bifurcation \cite{strogatz2024nonlinear}, stable dark solitons develop within the high loss regions \footnote{
The breaking of the translational symmetry induced by $V(x)$ induces a transcritical bifurcation \cite{strogatz2024nonlinear} which result in stable dark (bright) solitons forming on the high (low) loss regions, while unstable dark (bright) solitons form on the low (high) loss ones. The latter and unstable types are thus not discussed in this work.
} and they undergo an abrupt SN bifurcation when these LSs achieve a width comparable to the modulation. Past that SN, the outer switching waves detach from the potential region [panel $(iii)$] and the snaking consequently collapses around the MP. Annihilation of the moving fronts, due to spatial periodicity, eventually results in the formation of the so-coveted single and narrow bright soliton (extremely wide dark soliton)  \cite{xue2017microresonator}, see panels $(iv)-(v)$ in Fig.2(b), which stable existence spans over an unprecedented large interval \cite{parra2021influence,parra2017coexistence,parra2022quartic,ivars2024hybrid,zhang2023quintic}.

The above drastic unfolding of the bifurcation diagram in Fig.\ref{fig:1}(b) into those discussed in Figs.\ref{fig:1}(d) and \ref{fig:2}(a) introduce exciting prospects for real-time excitation and manipulation of LSs and the associated frequency combs, where pump detuning stands out as a powerful knob to tune combs at will. In particular, the unfolded bifurcation diagrams of the kinds shown in Figs.1(d) and 2(a) [see also Fig.4(a) below] enable the adiabatic and deterministic excitation of frequency combs in all-normal dispersion Kerr cavities. In the absence of a potential, $V(x)$, this type of cavities requires either dispersion management \cite{savchenkov2012kerr,kim2019turn,xue2019super,rebolledo2023platicon,sanyal2025nonlinear} or self-injection-locking \cite{jin2021hertz,lihachev2022platicon,kondratiev2023recent,li2025universal} to successfully excite solitons and combs. In the here-reported approach, modulation instability and mode crossing requirements are removed so that the device maintains an all-normal dispersion landscape. Figure \ref{fig:3}(a) shows, via direct time integrations of Eqs.\ref{eq:1}-\ref{eq:potential}, the dynamical transformations of the quasi-homogeneous states for the case in Fig.\ref{fig:1}(d) [cf. arrow therein with label \ref{fig:3}(a)] as detuning is continuously red-shifted in time [dotted line in bottom panel of Fig. 3(a)]. During this laser red-shift, the quasi-homogeneous states smoothly transit into the whole family of single dark solitons [$2.8\lesssim\delta\lesssim6$], followed by an abrupt switching [at the SN$_d$ point in Fig.\ref{fig:1}(d)] into the stable double bright LSs. We note that the system presumably prefers to decay from single dark LS into double bright LS instead of decaying to the bottom quasi-homogeneous state branch because the LS states are matched in the Full Width at Half minimum (FWHm), $w$, calculated in-between the central lobe of the modulation [solid line in bottom panel of Fig. 3(a)]. 

Similarly, red-shifting the pump along the top branch in Fig.\ref{fig:2} produces the smooth excitation of double dark LSs followed by an abrupt switching into the single bright solitons (not shown). Moreover, and in sharp contrast with usual red-shift approaches, an excitation path appears for the bright soliton by a blue detuning of the laser [cf. arrow in Fig.\ref{fig:2} with label \ref{fig:3}(b)]. Here, the bright soliton is abruptly excited from the quasi-homogeneous state as the blue detuning crosses the SN$_2$ point [see abrupt jump in the maximum amplitude of the field at the bottom panel $t\lesssim 12$ in \ref{fig:3}(b)]. Thereafter, red-shifting the pump ($t\gtrsim13$) results in smooth transformations over the bright LS branch.


Thus far, we have illustrated how the unfolding of collapsed homoclinic snaking opens new routes for adiabatically exciting dark LSs as well as in tuning their widths. Below, we discuss our final and most important result regarding the scope of such an unfolding mechanism. We recall [see above discussion about Fig.\ref{fig:1}(d)] that $q$ determined the number of cusp bifurcations and with it the number of SNs that we were able to unfold. Indeed, decreasing $q$ widens the potential and increases the number of SNs [of Fig.1(b)] that can be eliminated via cusp bifurcations, so that more and more states of the [originally] collapsed snaking may be continuously accessed. Strikingly, low $q\approx0.03$ values (or wide potentials, see $V(x)$ in Fig.\ref{fig:4}) further reshape the bifurcation diagram in Fig.\ref{fig:1}(d) into that shown in Fig.\ref{fig:4}(a), where all stable LSs of the system, including dark and narrow bright LSs, lie on the one same branch which smoothly emerges from the quasi-continuous background and feature no SNs nor unstable regions of any kind. The LSs and corresponding selected frequency combs are shown in Figs.\ref{fig:4}(b)-right and -left, respectively. Thus, we have uncovered the possibly simplest mechanism so far to excite LSs of virtually any width by simply red-shifting the cavity-laser detuning in a fixed cavity geometry. We emphasise that this result, product of a judicious choice of a non-Hermitian potential, could not be anticipated \textit{a priori} and constitutes the central finding of this work.


A particularly relevant aspect of the narrow bright LSs in normal regime is the fact that they display a prominent pump-to-comb efficiency, related to the energy effectively converted from the injected pump to the other lines of the comb. Efficiency may be evaluated in direct space from the spatial duty cycle \cite{xue2017microresonator}, which increases dramatically for bright LSs, or, similarly, from the Fourier space as the line contrast in the intra-cavity field $\eta=P_{k\neq0}/P$ where $P$ represents the power of the comb and $P_{k\neq0}$ the power excluding the central line. We found similar values of $\eta$ with both methods and the results are shown in Fig.\ref{fig:4}(a) for $q=0.03$ and in Fig.\ref{fig:4}(c) over the plane $\{\delta,q\}$. As observed, there exists an optimal $q\approx0.03$ which dramatically increases the efficiency of the adiabatically accessible combs up to values exceeding 90$\%$.

Localised non-Hermitian potentials offer a novel mechanism for the adiabatic and deterministic excitation of frequency combs in Kerr cavities with normal dispersion, without relying on mode crossings or modulational instability. These potentials trigger different cusp bifurcations that unfold the otherwise collapsed snaking bifurcation structure of solitons with normal GVD, placing all dark and bright solitons as states that smoothly emerge from the quasi-homogeneous state. The reported results establish a new paradigm for understanding the formation of dissipative localised states where the destabilisation of the cavity background is not involved. This unconventional scenario is central for the here-reported adiabatic \textcolor{black}{and continuous} excitation, which leads to atypical, yet deterministic, reconfiguration of frequency combs and access to highly efficient comb generation.

\begin{acknowledgments}
This work was partially supported by the Ministerio de Economía y Competitividad, funded by MCIN/AEI/10.13039/501100011033/FEDER, Agencia Estatal de Investigación, (PID2022-138280NB-I00),  Agència de Gestió d'Ajuts Universitaris i de Recerca (2021 SGR 01448), CEX2024-001490-S [MICIU/AEI/10.13039/501100011033], Fundació Cellex, Fundació Mir-Puig, and Generalitat de Catalunya through CERCA. C.M.E acknowledges funding from Agencia Estatal de Investigación ALLEGRO PID2021- 124618NB-C21 and INCEPTION PID2024-157370NB-I00, C.M.A acknowledges funding from the Ministerio de Universidades through the Beatriz Galindo program (BG22/00025).
\end{acknowledgments}

\nocite{} 

\bibliography{apssamp}

\end{document}